\documentclass[a4paper,aps,prl,preprintnumbers,twocolumn,amsmath,amssymb]{revtex4}
\usepackage{bbm}
\usepackage{amsmath}
\usepackage{verbatim}
\usepackage{graphicx}
\usepackage{epsfig}

\def\<{\langle}
\def\>{\rangle}

\newcommand{\be}{\begin{equation}}
\newcommand{\ee}{\end{equation}}
\newcommand{\bea}{\begin{eqnarray}}
\newcommand{\eea}{\end{eqnarray}}
\newcommand{\ii}{\mathbbm{1}}

\newcommand{\tr}[1]{{\rm tr}\left[#1\right]}

\begin{document}
\title{\textbf{Violation of the entropic area law for Fermions}}

\author{Michael M. Wolf}
\affiliation{\normalsize  Max-Planck-Institute for Quantum Optics,
 Hans-Kopfermann Str.1, 85748 Garching, Germany.}

\begin{abstract}
We investigate the scaling of the entanglement entropy in an
infinite translational invariant Fermionic system of any spatial
dimension. The states under consideration are ground states and
excitations of tight-binding Hamiltonians with arbitrary
interactions. We show that the entropy of a finite region
typically scales with the area of the surface times a logarithmic
correction. Thus, in contrast to analogous Bosonic systems, the
entropic area law is violated for Fermions.  The relation between
the entanglement entropy and the structure of the Fermi surface is
discussed, and it is proven, that the presented scaling law holds
whenever the Fermi surface is finite. This is in particular true
for all ground states of Hamiltonians with finite range
interactions.
\end{abstract}

\date{\today}

\maketitle

Entanglement is a phenomenon of common interest in the fields of
quantum information and condensed matter theory. It is an
essential resource for quantum information processing and
intimately connected with exciting quantum phenomena like
superconductivity, the fractional quantum Hall effect or quantum
phase transitions. Crucial to all these effects are quantum
correlations, i.e., the entanglement properties, of ground states.
These have recently attracted a lot of attention, leading to new
insight into quantum phase transitions and renormalization group
transformations \cite{rgflows} and triggering the development of
new powerful numerical algorithms \cite{numerics}.

A fundamental question in this field is concerned with the scaling
of the entropy---which is for pure states synonymous with the
entanglement. That is, given a ground state of a translational
invariant system, how does the entropy of a subsystem grow with
the size of the considered region?
 Originally, this question
appeared first in the context of black holes, where it is known
that the Bekenstein entropy \cite{beckenstein} is proportional to
the area of the horizon, which led to the famous conjecture now
known as the \emph{holographic principle} \cite{holo,holo2}. The
renewed interest, however, comes more from the investigation of
spin systems and quantum phase transitions. Moreover, the scaling
of the entropy is of particular interest concerning the choice of
the right ansatz-states in simulation algorithms.

In the last years, especially one-dimensional spin chains have
been  studied extensively and it is now believed that the entropy
diverges logarithmically with the size of a block if the system is
critical, and that it saturates at a finite value otherwise
\cite{Latorre}. For a number of models
\cite{Peschel,Korepin1,random2}, in particular those related to
conformal field theories in 1+1 dimensions \cite{Korepin2,CFT},
this could be shown analytically revealing a remarkable connection
between the entropy growth and the universality class of the
underlying theory. At the same time the diverging number of
relevant degrees of freedom provides a simple understanding of the
failure of DMRG methods for critical spin-chains.

For several spatial dimensions a suggested entropic area law
\cite{harm1}  could recently be proven \cite{harm2} for the case
of a lattice of quantum harmonic oscillators (quasi-free Bosons),
where again the entropy grows asymptotically proportional to the
surface. On heuristic grounds this can be understood from the fact
the system is non-critical: an energy gap gives rise to a finite
correlation length, which in turn defines the scale on which modes
inside the subsystem are correlated with the exterior. Although a
general area law for gapped lattice systems has not been proven so
far, the case of quasi-free Bosons is often considered as
paradigmatic. In fact, recently developed simulation algorithms
based on ansatz-states exhibiting the presumed entropy scaling are
highly promising \cite{numerics}.

The fact that in some 1-$d$ systems a vanishing energy gap leading
to a diverging correlation length results in the logarithmically
diverging entanglement entropy inevitably raises the question
about the behavior of {\it gapless} systems in {\it more} than one
dimension.

The present paper is devoted to the study of the entanglement
entropy in gapless Fermionic systems of arbitrary spatial
dimensions. We establish a relation between the structure of the
Fermi sea and the scaling of the entropy and prove that a finite
non-zero Fermi surface implies that the entanglement grows
proportional to the surface of the subsystem times a logarithmic
correction, i.e.,\be S\sim L^{d-1}\log L\;,\ee if the system under
consideration is a $d$-dimensional cube with edge length $L$.
Thus, in contrast to analogous Bosonic systems the entropic area
law is violated for Fermions.

Before we start to prove this result a brief discussion of the
notion of locality---necessary for the concept of
entanglement---is in order. In spin systems as well as in the
Bosonic case of harmonic oscillators the tensor product structure
of the underlying Hilbert space naturally leads to an unambiguous
notion of locality. In the absence of such a tensor product
structure either in general quantum field theory settings
\cite{VW} or in the present case of Fermions \cite{BK}, one has to
identify commuting subalgebras of observables
 and assign them to different parties. In our case the
 relevant algebra is the one spanned by the Fermionic creation and
 annihilation operators $c_j^\dagger$ and $c_j$ satisfying the usual anti-commutation relations.
 We assign the modes with $j=1,\ldots,n$ to one party $\cal A$
(the
 interior) and the other modes $j=n+1,n+2,\ldots$ to the other
 party $\cal B$ (the exterior). Then parity conservation or even stronger, particle
 conservation \cite{Wick}, leads to a superselection rule, which implies that
 all physical operators acting on $\cal A$ commute with those
 acting on $\cal B$, leading to well-defined notions of locality and entanglement.

Let us now introduce the prerequisits for the proof. Consider a
number preserving quadratic Hamiltonian \be\label{H}
\hat{H}=\sum_{\alpha,\beta\in\mathbb{Z}^d} T_{\alpha,\beta}
c_\alpha^\dagger c_\beta\;,\quad T=T^\dagger\;,\ee describing
Fermions on a $d$-dimensional cubic lattice, so that each
component of the vector indices $\alpha,\beta$ corresponds to one
spatial dimension. Translation symmetry is reflected by the fact
that $T$ is a Toeplitz operator, i.e.,
$T_{\alpha,\beta}=T_{\alpha-\beta}$ depends only on the distance
between two lattice points. The Hamiltonian (\ref{H}) is
diagonalized by a Fourier transform leading to the dispersion
relation \be \epsilon(k)=\sum_{\alpha\in\mathbb{Z}^d} T_\alpha\;
e^{-i k\cdot\alpha}\;,\quad k\in[-\pi,\pi]^d\;. \ee All thermal
and excited states of the Hamiltonian $\hat{H}$ are Fermionic
Gaussian state \cite{SB04}, which are completely characterized by
their correlation matrix
\be\gamma_{\alpha\beta}=\delta_{\alpha\beta}-2\tr{\rho
c_\alpha^\dagger c_\beta}\;.\ee The correlation matrix describes a
pure state iff $\gamma^2=\ii$, so that all eigenvalues of $\gamma$
are $\pm 1$, and the ground state correlation matrix is given by
$\gamma=\frac{T}{|T|}$. Ground states for different Fermion
densities are then obtained by adding a chemical potential, i.e.,
replacing $T$ by $T+\mu\ii$. If we characterize the Fermi sea by
the corresponding indicator function $\theta(k)\in\{0,1\}$, then
the respective correlation matrix is given by \cite{Fermitheta}
\be\label{gamma}
\gamma_{\alpha\beta}=\frac1{(2\pi)^d}\int_{-\pi}^{\pi}
dk_1\ldots\!\int_{-\pi}^{\pi} dk_d \big[1-2\theta(k)\big]\; e^{i
k\cdot(\alpha-\beta)}\;.\ee Note that this characterizes not only
ground states but pure Gaussian states in their most general form
(as long as they obey particle conservation and translation
symmetry).

The state of a subsystem, e.g., a cube with edge length $L$, is
described by the corresponding $L^d\times L^d$ sub-matrix of
$\gamma$, which we will denote by $\tilde\gamma$. This subsystem
can be decomposed into normal modes by a canonical transformation
from $U(L^d)$ such that the state of each normal mode has a Fock
space representation of the form \be
\frac{1-\lambda_j}2\;|1\rangle\langle1|
+\frac{1+\lambda_j}2\;|0\rangle\langle0|\;, \ee where the
$\lambda_j$ are the eigenvalues of $\tilde\gamma$. The entropy of
the subsystem can then be expressed as \bea
S(\tilde{\gamma})&=&\sum_{j=1}^{L^d}\; h(\lambda_j)\;,\\
h(x)&=& -\frac{1+x}2\log\frac{1+x}2-\frac{1-x}2\log\frac{1-x}2\;.
\eea Since a direct computation of $S(\tilde{\gamma})$ via the
diagonalization of $\tilde\gamma$ is yet highly non-trivial in the
simplest one-dimensional case with nearest neighbor interaction
\cite{Korepin1}, one relies in general on finding good bounds on
the entropy. We will use quadratic bounds on $h(x)$ of the form
$f(x)=a(1-x^2)+b$ \cite{Fannes}. The best lower bound is given by
$a=1,\;b=0$ leading to \be S(\tilde{\gamma})\geq
\tr{\ii-\tilde{\gamma}^2}\;.\ee The set of tight quadratic upper
bounds can be parameterized by the point $x_0\in[0,1)$ for which
$f(x_0)=h(x_0)$ become tangent \cite{x0}. We will couple this
bound to the block size $L$ via $x_0=1-1/g(L)$, where $g(L)=L/\log
L$. Straight forward but lengthy calculations show then that the
entropy as a function of $L$ is asymptotically upper bounded
\cite{Landau} by \be S(\tilde{\gamma})\leq
O\big(\tr{\ii-\tilde{\gamma}^2}\log
g(L)\big)\;.\label{upperbound}\ee Hence, together with the lower
bound this means that $\tr{\ii-\tilde{\gamma}^2}$ essentially
determines the asymptotic scaling of the entropy.

The necessity of coupling the upper bound to $L$ can easily be
understood physically, when one recalls that there is always a
choice of the local basis in which each normal mode inside the
block is only correlated with at most one mode outside
\cite{BR04}. With increasing block size, more and more modes
inside lose their correlations with the exterior, such that the
number of nearly pure normal modes dominates more and more. Since
the corresponding eigenvalues are $\lambda_j\simeq\pm 1$, the
point $x_0$, for which the bound is tight, should tend to 1 as
$L\rightarrow\infty$. Setting $x_0=1$ right from the beginning is,
however, not possible since the derivative of $h(x)$ at this point
diverges.

Let us now investigate the scaling of $\tr{\ii-\tilde{\gamma}^2}$.
To this end we introduce the positive Fej\'er kernel \cite{DM} \be
F_L(x) = \sum_{\alpha,\beta\in\mathbb{Z}_L} e^{i
x(\alpha-\beta)}=\frac{\cos(Lx)-1}{\cos(x)-1}\;,\ee and we will
abbreviate $\prod_{i=1}^dF_L(k_i)$ by $F_L(k)$. Then following
Eq.(\ref{gamma}) we have  \bea \tr{\ii-\tilde{\gamma}^2} &=&
\!\frac4{(2\pi)^{2d}}\!\int {dk dk'}\;
\theta(k)\big[1-{\theta}(k')\big] F_L(k-k')\nonumber\\
&=& \frac4{(2\pi)^{2d}}\int dq \;\Xi(q)\; F_L(q)\;,\label{XD}\\
\Xi(q)&=&\int dk\; \theta(k)\big[1-\theta(q+k)\big]\;.\eea To
further evaluate Eq.(\ref{XD}) we have to exploit the fact that
with increasing $L$ the Fej\'er kernel $F_L(x)$ becomes more and
more concentrated around $x=0$. In fact, $F_L(0)=L^2$,
$\int_{-\pi}^\pi dx F_L(x)=2\pi L$ and for all $\epsilon>0$ there
exists a finite constant $c_{\epsilon}$ such that
\be\int_{[-\pi,\pi]^d} \!\!dq \;\Xi(q)\; F_L(q) \leq\;
c_\epsilon+\! \int_{[-\epsilon,\epsilon]^d}\!\! dq \;\Xi(q)\;
F_L(q) \;.\label{epsilonint}\ee The crucial point here is that
$c_\epsilon$ does not depend on $L$. Hence, the asymptotic scaling
of the entropy depends only on the behavior of the function
$\Xi(q)$ in an $\epsilon$-neighborhood of the origin.

The function $\Xi(q)$ has a very intuitive interpretation: it is
the volume of the part of the Fermi sea in $k$-space, which is no
longer covered if we shift the Fermi surface by a vector $q$ (see
Fig.1). So what is the behavior of $\Xi(q)$ near the origin?
Obviously, $\Xi(0)=0$ and $\Xi(q)\geq 0$. Moreover, $\Xi$ will
typically not be differentiable at $q=0$, but it will rather have
the structure of a pointed cone.

Let us assume that the Fermi sea is a set of non-zero measure with
a finite non-zero surface. This means in particular, that almost
all points with $\theta(k)=1$ are interior points of the Fermi sea
and it implies that $\Xi$ is a continuous function in $k$-space
\cite{continuity}. In fact, an infinite boundary could lead to a
discontinuity of $\Xi$ at the origin.
 This restriction excludes both,
trivial cases (zero entropy due to zero surface) and exotic cases
(fractal or Cantor set like Fermi seas).  For all other cases it
enables us to bound $\Xi(q)$ in a neighborhood of the origin by
pointed cones in the following manner: consider the surface of the
closed interior of the Fermi sea in one unit cell of the
reciprocal lattice. Let $s(q)$ be the area of the projection of
this surface onto the hyper-plane with normal vector $q$, where we
account for each front of the surface. If, for example, the Fermi
sea consist out of two disjoint three-dimensional spheres with
radius $r$, then $s(q)=2\pi r^2$ in every direction. Since
$\Xi(q)$ is the volume in which the Fermi sea changes upon
shifting it by $q$, we have in an $\epsilon$-neighborhood of the
origin that $\Xi(q)$ is given by $ s(q) ||q||_2$. Using the fact
that the Fermi surface is assumed to be finite we know that $s(q)$
is bounded from above by a finite constant $s^+$. Let us assume
for the moment that there exists a non-zero lower bound $s^-$ as
well. Then we can bound the integral in Eq.(\ref{epsilonint}) by
replacing $\Xi(q)$ with $s^\pm ||q||_2$. Exploiting further that
$||q||_1\geq||q||_2\geq||q||_1/\sqrt{d}$ and that $F_L$ is
symmetric, leads to upper and lower bounds which are up to a
finite constant given by \bea\label{isum} 2^d s^\pm\!\!
\int_{[0,\pi]^d}\!\!\!\!dq F_L(q) ||q||_1 &=& 2^d
s^\pm\sum_{i=1}^d\int_{[0,\pi]^d}\!\!\!\! dq F_L(q)q_i\\&=&2ds^\pm
(2 \pi L)^{d-1}\!\!\int_0^\pi \!\!\!dxF_L(x)x\;.\nonumber\eea The
remaining integral is the Fej\'er sum of a linear function, which
can be evaluated  \cite{Landau} to \bea \int_0^\pi \!\!\!dx
F_L(x)x&=&2\Big(1+c_\gamma+\ln
2+\psi(L)+O(L^{\!-1})\Big)\nonumber\\&=&2\ln L +O(1)\;,\eea where
$c_\gamma\simeq 0.577$ is Euler's constant and $\psi$ denotes the
digamma function.

We still have to discuss the case $\inf_q s(q)=0$. In this case we
have to use different linear bounds for different directions in
Eq.(\ref{isum}). Since $s(q)$ will be larger than some $s^->0$ at
least in one direction, we are in the end led to the same type of
integral and thus to the same asymptotic scaling.

\begin{figure}[t]
\begin{center}
\epsfig{file=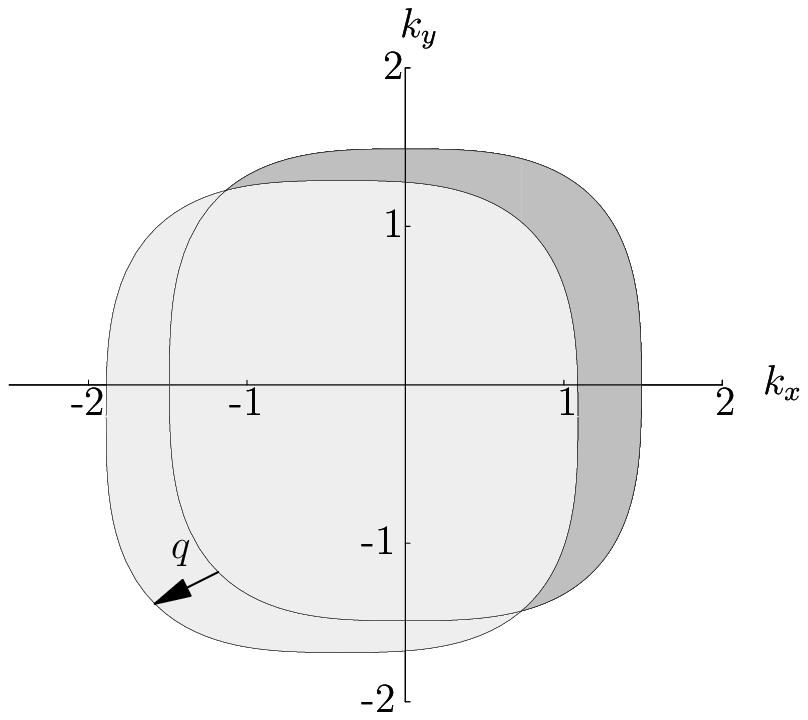,angle=0,width=0.55\linewidth}
\end{center}
\caption {Consider the Fermi surface and shift it by a vector $q$
in $k$-space. $\Xi(q)$ is then given by the (dark gray) area of
the Fermi sea which is no longer covered by its translation. The
scaling of the entanglement entropy depends only on the behavior
of $\Xi$ in the vicinity of $q=0$.} \label{figure}
\end{figure}

 Putting it all together we have indeed that
$\tr{\ii-\tilde{\gamma}^2}$ scales as $L^{(d-1)}\ln L$ since all
the involved constants are finite and depend not on $L$ but merely
on the structure of the Fermi sea. The above argumentation holds
under the assumption that the Fermi surface is not too exotic.
However, if the interactions are finite in range, then the Fermi
surface of the ground state is differentiable to infinite order
and it is in particular finite.
In general, however, one has to
check whether or not the structure of the Fermi sea gives rise to
an infinite slope or a discontinuity of $\Xi(q)$ at the origin.

The existence of Fermi surfaces leading to a scaling of the
entanglement entropy which surpasses the above law, can easily be
understood: consider a Fermi sea given by a checkerboard with
squares of edge length $l$. Then shifting the Fermi surface by $l$
along any lattice axis yields $\Xi=2\pi^2$ such that a naive limit
$l\rightarrow 0$ would indeed give rise to a $L^d$ scaling of the
entropy. Needless to say that the checkerboard does not have a
well-defined limit---however, following the same idea, more
sophisticated Cantor set like constructions will do the same job
without any caveat. In fact, for $d=1$ such states were
constructed in \cite{Fannes}.

Remarkably, fractal or Cantor set like structures are known to
appear in tight binding models. The most prominent example is the
Azbel-Hofstadter Hamiltonian \cite{AB} with non-integer flux,
leading to the famous Hofstadter butterfly for the spectrum. Since
the interaction matrix $T_{kl}=\exp i \int_k^l A(s) ds$ (with $A$
being the vector potential) is quasi-periodic and not
translational invariant, this case is, however, not directly
covered by the above argumentation. The question, which physically
interesting translational invariant Hamiltonians give rise to a
violation of the above scaling law via a fractional Fermi sea,
remains an interesting problem for future research.

In conclusion we derived a method of relating the structure of the
Fermi sea in tight-binding models to the scaling of the
entanglement entropy.  For every finite non-zero Fermi surface we
proved the  violation of a strict area law (as it is assumed for
non-critical systems \cite{harm1,harm2,Zanardi}) by a logarithmic
correction, i.e., \be c_- L^{d-1} \log L\; \leq \; S\; \leq\; c_+
L^{d-1} (\log L)^2\;, \ee with constants $c_\pm$ depending only on
the Fermi sea. By the strong sub-additivity of the entropy the
same scaling behavior holds true also for other regions, e.g.
spheres, as long as they can be nested into two cubes of edge
lengths $L$ and $c L$ with $c$ independent of $L$. The additional
$\log L$ in the upper bound is presumably an artefact (cf.
\cite{random2,other}) coming from the incompatibility of tight
quadratic upper bounds with the binary entropy function at $\pm
1$.

Note finally, that the derived result can be applied to spin
models in one dimension \cite{random2,Fannes}. In this case a
Jordan-Wigner transformation maps Fermionic operators onto Pauli
spin operators such that every  tight-binding Hamiltonian with
nearest neighbor interactions in Eq.(\ref{H}) is then mapped onto
a spin Hamiltonian of the form \bea \hat{H}_\sigma &=&\sum_\alpha
h_0
\sigma_\alpha^z\;+\;h_1\big(\sigma_\alpha^x\sigma_{\alpha+1}^x+\sigma_\alpha^y\sigma_{\alpha+1}^y\big)\label{eq1}\\
\label{eq2} &&\ \quad\qquad\quad +\;h_2
\big(\sigma_\alpha^x\sigma_{\alpha+1}^y-\sigma_\alpha^y\sigma_{\alpha+1}^x\big)\;,\eea
with some couplings $h_i$. Conversely, every such Hamiltonian is
covered by Eq.(\ref{H}), and  we are in general allowed to add
arbitrary interaction terms differing from those in
Eqs.(\ref{eq1},\ref{eq2}) by a sequence of $\sigma^z$s in between
every two Pauli operators. For higher dimensions, however, an
analogous construction fails, since then Jordan-Wigner
transformations do no longer preserve locality. \vspace{2pt}\\
\emph{Acknowledgements} The author thanks T. Cubitt, J. Eisert, D.
Porras and I. Cirac for valuable discussions and D. Gioev and I.
Klich, who addressed the same question independently \cite{other},
for bringing Ref. \cite{Fannes} to his attention.

\newpage


\begin{thebibliography}{99}
\bibitem{holo2} Despite the motivation for \cite{harm1} the
relation to black hole entropies and the holographic principle is
not entirely clear \cite{holo}. After all, gravity does not enter
and the size of the Hilbert space is a priori unrelated to the
entanglement entropy.
\bibitem{Fermitheta} $\theta(k)$ is a function in $k$-space which is
one if the wave vector $k$ is inside the Fermi sea, and zero
otherwise. For ground states
$\theta(k)=(1-\epsilon(k)/|\epsilon(k)|)/2$.
\bibitem{x0} The upper bound $f(x)=a(1-x^2)+b\geq h(x)$, which is tight at $x_0$, is found by solving $f(x_0)=h(x_0)$ and $f'(x_0)=h'(x_0)$. This gives
{\footnotesize \bea a&=& \big[\log(1+x_0)-\log(1-x_0)\big]/[4 x_0],\nonumber\\
b&=&1+\big[(1-x_0)^2\log(1-x_0)-(1+x_0)^2\log(1+x_0)\big]/[4x_0].\nonumber\eea}
\bibitem{Landau} {Here Landau's asymptotic notation is used:
 $\eta=O(\nu)$
means that $\eta<C\nu$ for some constant $C$.}
\bibitem{continuity} In fact, $\Xi(q)-\Xi(q+\delta q)\leq ||\delta q||
S_F$, where $S_F$ is the area of the Fermi surface.
\end{thebibliography}
\end{document}